\documentclass[doublecol]{epl2}

\usepackage{mathptm}    %clears up some problems for generating PDF file
\usepackage{dcolumn}                    % Align table columns on decimal point
\usepackage{bm}                         % bold math
\usepackage{graphicx}

\newcommand {\ba} {\begin{eqnarray}}
\newcommand {\ea} {\end{eqnarray}}

\usepackage{times}

\title{Superfluid density of the $\pm$s-wave state for the iron-based superconductors}

\author{Yunkyu Bang\inst{1,2}}
\shortauthor{Yunkyu Bang}

\institute{
  \inst{1} Department of Physics, Chonnam National University,
Kwangju 500-757, Korea \\
  \inst{2} Asia Pacific Center for Theoretical Physics, Pohang 790-784, Korea
}


\abstract {We study the superfluid density of the $\pm$s-wave
state of the minimal two band model for the Fe-based
superconductors and its evolution with impurity concentration. We
show that the impurity scattering of the strong coupling limit
induces the selfenergy of a generic form $Im \Sigma_{imp} (\omega)
\approx i \gamma + i \beta \omega$ beyond a critical impurity
concentration $\Gamma_{imp} > \Gamma_{crit}$. This form of $Im
\Sigma_{imp} (\omega)$ causes the temperature dependence of the
superfluid density $[\rho_{s}(T) - \rho_{s}(0)] \approx - \gamma
T^2 -\beta T^3$. Combining with the full gap behavior of
$\rho_{s}(T)$ for lower impurity concentration $\Gamma_{imp} <
\Gamma_{crit}$, the $\pm$s-wave state produces a continuous
evolution of $\Delta \lambda(T)$: exponentially flat $\rightarrow
T^3 \rightarrow T^2$ with increasing impurity concentration that
is consistent with the measurements of the Fe pnictide
superconductors such as $M$-1111 ($M$=La,Nd,Sm,Pr) and Ba-122 with
various dopings, except LaFePO which shows  $\Delta \lambda(T)
\propto T^{1.2}$ at low temperatures by recent experiment. Our
results also demonstrate that the density of states (DOS) measured
by thermodynamic properties and the DOS measured by transport
properties can in general be different.}

\begin{document}

\maketitle

\section{Introduction}
The recent discovery of Fe-based superconducting compounds
\cite{Kamihara,org1,org2}, has greatly spurred the research of
unconventional superconductors. With a discovery of new
superconducting (SC) material, the most impelling question is to
determine the SC gap symmetry. However, despite intensive
experimental efforts, the pairing symmetry of this new class of SC
materials is not completely settled.

Tunneling spectroscopy of Ref.\cite{tunneling-d1,tunneling-d2},
photoemission spectroscopy of Ref.\cite{photoemission-d}, and NMR
nuclear-spin-lattice relaxation rate measurements
\cite{T1a,T1b,T1c,T1d,T1e} seem to indicate a nodal
superconductors (SCs). On the other hand, tunneling spectroscopy
of Ref.\cite{tunneling-s}, angle resolved photoemission
spectroscopy (ARPES) measurements of
Ref.\cite{photoemission-s1,photoemission-s2,photoemission-s3},
specific heat measurement \cite{C(T)} all support a fully opened
s-wave type gap. As to the penetration depth measurements, early
measurements of $M$-1111 ($M$=Pr, Nd, Sm)
\cite{1111a,1111b,1111c}, and (Ba,K)Fe$_2$As$_2$ \cite{Ba-122}
showed an exponentially flat behavior at low temperatures
supporting a s-wave type SC. However, recent reports of
Ba(Fe,Co)$_2$As$_2$ \cite{BaCo-122-a,BaCo-122-b} shows a power law
of $T^{2-2.5}$ and more recently measurement of LaFePO
\cite{LaFePO} shows a near $T$-linear temperature dependence.

On the theoretical side, it is almost agreed on that non-phononic
fluctuation is the most probable pairing interaction and several
theoretical proposals for the possible pairing symmetries of Fe
pnictide SCs were put forward
\cite{Mazin,other-theory1,other-theory2,other-theory3,Bang-model}.
However, recent advance of theoretical studies by several groups
\cite{Bang-imp,theory-T1a,theory-T1b,theory-T1c} have provided
convincing evidences that the $\pm$s-wave SC state is the most
promising candidate for the true pairing state of Fe pnictide SCs.
Therefore, we will consider only the $\pm$s-wave SC state in this
paper.

Having s-wave gaps on each bands but with opposite signs, the
$\pm$s-wave SC state would display standard features of a s-wave
SC for various SC properties at its pure state, for example, the
exponentially flat temperature dependence of the penetration
depth, which is consistent with some of experiments
\cite{1111a,1111b,1111c,Ba-122}. Another hallmark of a s-wave
pairing state is the Hebel-Slichter peak and exponential drop of
the nuclear- spin-lattice relaxation rate $1/T_1$, which is
absolutely inconsistent with the experimental reports of all Fe
pnictide SC compounds \cite{T1a,T1b,T1c,T1d,T1e}. It was quickly
pointed out by several groups
\cite{Bang-model,Bang-imp,theory-T1a,theory-T1b,theory-T1c} that
the opposite signs of the order parameters (OP) in the $\pm$s-wave
SC state with a help of impurity scattering could reduce the
Hebel-Slichter peak and also mitigate the exponential temperature
dependence into a power law temperature dependence.
In particular, the Ref.\cite{Bang-imp} showed the special
importance of the strong coupling (unitary limit) impurity
scattering in the $\pm$s-wave SC state. The source of strong
coupling impurities is also naturally expected when there are
detects or vacancies on Fe sites, of which the concentration
should be also very small (order of a few percents) unless the
superconductivity itself is substantially or completely
destructed.

To sketch briefly the previous work \cite{Bang-imp}, impurity
scattering of the strong coupling limit in the $\pm$s-wave SC
state induces an impurity bound state inside the SC gap, which is
however off-centered from zero energy due to the generic property
of the $\pm$s-wave state, namely, {\it the unequal sizes of the
s-wave OPs with opposite signs on different bands}. This impurity
bound state drastically modifies the fully gapped DOS of the pure
$\pm$s-wave state into a V-shaped DOS as in a d-wave SC.
This V-shaped DOS is not only consistent with the DOS directly
measured by tunneling spectroscopy
\cite{tunneling-d1,tunneling-d2} but also conform with the
isotropic gap measured by ARPES
\cite{photoemission-s1,photoemission-s2,photoemission-s3}; a
d-wave state would give a V-shaped DOS but at the same time should
show a strong anisotropy of the gap in the ARPES experiments.
Furthermore, it provides a clear physical explanation for the
origin of the $T^{\alpha} (\alpha \approx 3)$ power law of NMR
$1/T_1$ measured for all Fe pnictide SCs
\cite{T1a,T1b,T1c,T1d,T1e}. This is an excellent theoretical
success in that a noble interplay between the strong coupling
impurities and the $\pm$s-wave gaps provides coherent resolutions
to the several conflicting experimental observations with only one
adjustable parameter, i.e., the impurity concentration.

In this paper, we extend the previous work \cite{Bang-imp} to
study the superfluid density and penetration depth in order to
complete the consistency test of the $\pm$s-wave state as the
ground state of Fe pnictide SCs, except LaFePO \cite{LaFePO}. The
main results of this paper is that the temperature dependence of
$\lambda(T)$ of the $\pm$s-wave state evolves systematically with
increasing impurity concentration: from a exponentially flat
behavior for pure sample to a $T^3$ behavior at a critical
impurity concentration, and finally to a $T^2$ one beyond the
critical impurity concentration.
These temperature dependencies are in agreement with the data of
$M$-1111 ($M$=Pr, Nd, Sm) \cite{1111a,1111b,1111c} and
(Ba,K)Fe$_2$As$_2$ \cite{Ba-122} (exponentially flat), and
Ba(Fe,Co)$_2$As$_2$ \cite{BaCo-122-a,BaCo-122-b} (power law $
\propto T^{2-2.5}$). However, recent report of $\Delta \lambda(T)
\propto T^{1.2}$ for LaFePO \cite{LaFePO} is not compatible with
our scenario. Possible explanation for this exception is that this
compound may have a different pairing symmetry than the other Fe
pnictide SC compounds \cite{Bang-model}.  We will not pursue this
issue in the current paper.

\section{Model and Formalism} For the minimal two band model of the
$\pm$s-wave pairing state, we assume two s-wave OPs $\Delta_h$ and
$\Delta_e$ on the two representative bands of the Fe pnictide
materials: one hole band around $\Gamma$ point and one electron
band around $M$ point in the reduced Brillouin Zone. $\Delta_h$
and $\Delta_e$ have opposite signs and their magnitude are
different in general. Impurity scattering will renormalize the
energy ($\omega$) and the OPs ($\Delta_{h,e}$) through the
selfenergy corrections: normal selfenergy and anomalous
selfenergy, respectively. These impurity induced selfenergies are
calculated by the $\mathcal{T}$-matrix method
\cite{T-mtx-1,T-mtx-2}, suitably generalized for the $\pm$s-wave
pairing model \cite{Bang-imp}, as follows.

\ba \tilde{\omega}_n =\omega_n + \Sigma^{0}
 _h(\omega_n) + \Sigma^{0} _e(\omega_n), \label{eq.1} \\
\tilde{\Delta}_{h,e} = \Delta_{h,e} + \Sigma^1 _{h} (\omega_n) +
\Sigma^1 _{e} (\omega_n), \label{eq.2} \\
\Sigma_{h,e} ^{0,1} (\omega_n)  = \Gamma \cdot \mathcal{T}^{0,1}
_{h,e} (\omega_n), ~~ \Gamma= \frac{n_{imp}}{\pi N_{tot}},
\label{eq.3} \ea

\noindent where $\omega_n= T \pi (2n +1)$ is the Matsubara
frequency, $n_{imp}$ the impurity concentration, and
$N_{tot}=N_h(0) +N_e(0)$ is the total DOS. The
$\mathcal{T}$-matrices $\mathcal{T}^{0,1}$ are the Pauli matrices
$\tau^{0,1}$ components in the Nambu space and defined as follows.

\ba
\mathcal{T}^{i} _{a} (\omega_n) &=& \frac{G^{i} _{a} (\omega_n)}{D} ~~~~~(i=0,1; ~~a=h,e), \label{eq.4}\\
D &=& c^2 +[G^0 _h + G^0 _e]^2 + [G^1 _h + G^1 _e]^2, \label{eq.5}\\
G^0 _a (\omega_n) &=& \frac{N_a}{N_{tot}} \left\langle
\frac{\tilde{\omega}_n}
{\sqrt{\tilde{\omega}_n^2 + \tilde{\Delta}_{a} ^2 (k) }} \right\rangle, \label{eq.6}\\
G^1 _a (\omega_n) &=& \frac{N_a}{N_{tot}} \left\langle
\frac{\tilde{\Delta}_{a}} {\sqrt{\tilde{\omega}_n^2 +
\tilde{\Delta}_{a} ^2 (k) }}  \right\rangle , \label{eq.7} \ea

\noindent where $c=\cot \delta_0$ is a convenient measure of
scattering strength, with $c=0$ for the unitary limit and $c > 1$
for the Born limit scattering. $\langle ...\rangle$ denotes the
Fermi surface average.

With the self-consistently calculated $\tilde{\omega}_n$
(Eq.(\ref{eq.1})) and $\tilde{\Delta}_{h,e}$ (Eq.(\ref{eq.2})), we
can calculate all physical quantities of the model. For small
amount of impurity concentration ($\Gamma/\Delta_e \ll 1$) of the
strong scatterers ($c=0$), which we believe the case of the
relevant experiments, the $T_c$ suppression is small
\cite{Bang-imp} and the renormalization of the OPs
$\tilde{\Delta}_{h,e}$ is also marginal. However, the
renormalization of $\tilde{\omega}_n$ is strong even with a small
amount of unitary scatterers and develops an off-centered impurity
bound state inside the SC gap. This unusual low energy impurity
band modifies the fully gapped DOS of the pure state into a
V-shape DOS and becomes the origin of the power law ($\propto
T^3$) of the NMR relaxation rate $1/T_1$ of Fe pnitide SCs
\cite{T1a,T1b,T1c,T1d,T1e}. In this paper, we study how this
modified DOS of the $\pm$s-wave state with impurities affect the
superfluid density and penetration depth.

Avoiding unnecessary repeat, we begin with showing previous
calculations of the DOS $N_{tot}(\omega)$ and the corresponding
normal selfenergy $Im \Sigma^0 _{tot} (\omega) = Im \Sigma^0 _h +
Im \Sigma^0 _e$ for various impurity concentrations in
Fig.\ref{fig1}(a) and Fig.\ref{fig1}(b). These results were
calculated with the parameters $|\Delta_e| / |\Delta_h| = 2.5$
with $N_h (0)/N_e (0) = 2.6$, which were obtained with the
realistic band structure parameters \cite{Bang-imp}. The physical
results of the present paper, however, are independent of the
specific choice of theses parameters. All energy scales are
normalized by $|\Delta_e|$ in this paper. Fig.\ref{fig1}(a) shows
the total DOS with different impurity concentrations $\Gamma /
\Delta_e= 0.0, 0.01, 0.04, 0.08$ of the unitary scatterer (c=0),
and Fig.\ref{fig1}(b) shows the corresponding impurity induced
selfenergy $Im \Sigma^0 _{tot} (\omega) = Im \Sigma^0 _h + Im
\Sigma^0 _e$.

Figure \ref{fig1}(a) shows systematic evolution of how the fully
opened gap of pure state is filled with impurity states; the
pattern of filling is very unusual and the $\Gamma / \Delta_e=
0.04$ case displays a perfect V-shape DOS down to zero energy
similar to a pure d-wave SC DOS. The origin of this behavior is
easily seen in Fig.\ref{fig1}(b); the impurity bound state is
never formed at zero energy but away from it because of the
incomplete cancellation of $[G^1 _h + G^1 _e]$ in $D$
(Eq.(\ref{eq.5})), the denominator of the $\mathcal{T}$-matrices
$\mathcal{T}^{0,1}$. Were the sizes of $N_h |\Delta_h |$ and $N_e
|\Delta_e |$ equal, the FS averaged anomalous Green's functions
$[G^1 _h + G^1 _e]$ would vanish and then the impurity bound state
would form at zero energy; this condition is intrinsically
satisfied in the case of d-wave SC state. In generic $\pm$s-wave
state, therefore, {\it the full gap around $\omega=0$ is
protected} until this off-centered impurity band spills over to
the zero energy with increasing impurity concentration. When it
touches the zero energy limit, the $\pm$s-gap state will
thermodynamically behave like a pure d-wave SC, and this happens
with the critical impurity concentration $\Gamma_{crit}$ ($=0.04
\Delta_e$ for our specific model parameters). Increasing impurity
concentration beyond $\Gamma_{crit}$, the DOS still keeps the
V-shape but now $N_{tot} (\omega=0)$ obtains a finite value (see
the $\Gamma=0.08 \Delta_e$ case in Fig.\ref{fig1}(a)).

\section{Superfluid density and penetration depth} With the above
calculated $\tilde{\omega}_n$ and $\tilde{\Delta}_{h,e}$, we can
study the temperature dependence of the superfluid density along
the Fe-plane, with the following kernel.

\begin{equation}
K(T)= \frac{e^2}{c} \sum_{a=h,e} N_{a} 2 \pi T \sum_n \left
\langle v_{a \|} ^2 {\rm Re}
\frac{\tilde{\Delta}_{a}^2}{(\tilde{\omega}_n ^2 +
\tilde{\Delta}_{a}^2)^{3/2}} \right \rangle. \label{eq.8}
\end{equation}

\begin{figure}
\noindent
\onefigure[width=80mm]{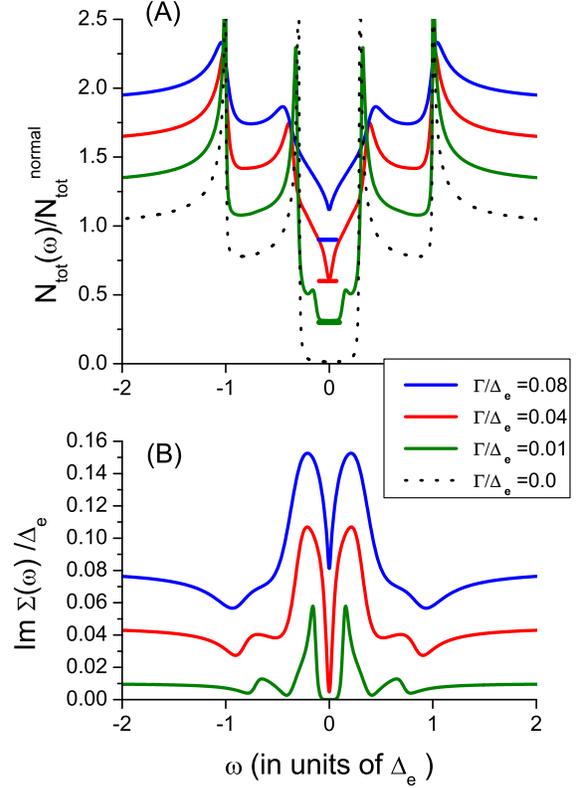}
\caption{(Color online) (a) Normalized DOS $N_{tot}(\omega)$ for
different impurity concentrations, $\Gamma / \Delta_e= 0.0, 0.01,
0.04, 0.08$. Thin dotted line is of the pure state for comparison
and other lines are offset for clarity (the zero baselines of the
offset are marked by the narrow horizontal bars of the
corresponding colors). (b) Impurity induced selfenergies $Im
\Sigma^0 _{tot} (\omega) = Im \Sigma^0 _h + Im \Sigma^0 _e$ with
the same parameters as in (a). These curves are not offset.}
\label{fig1}
\end{figure}

\begin{figure}
\noindent
\onefigure[width=80mm]{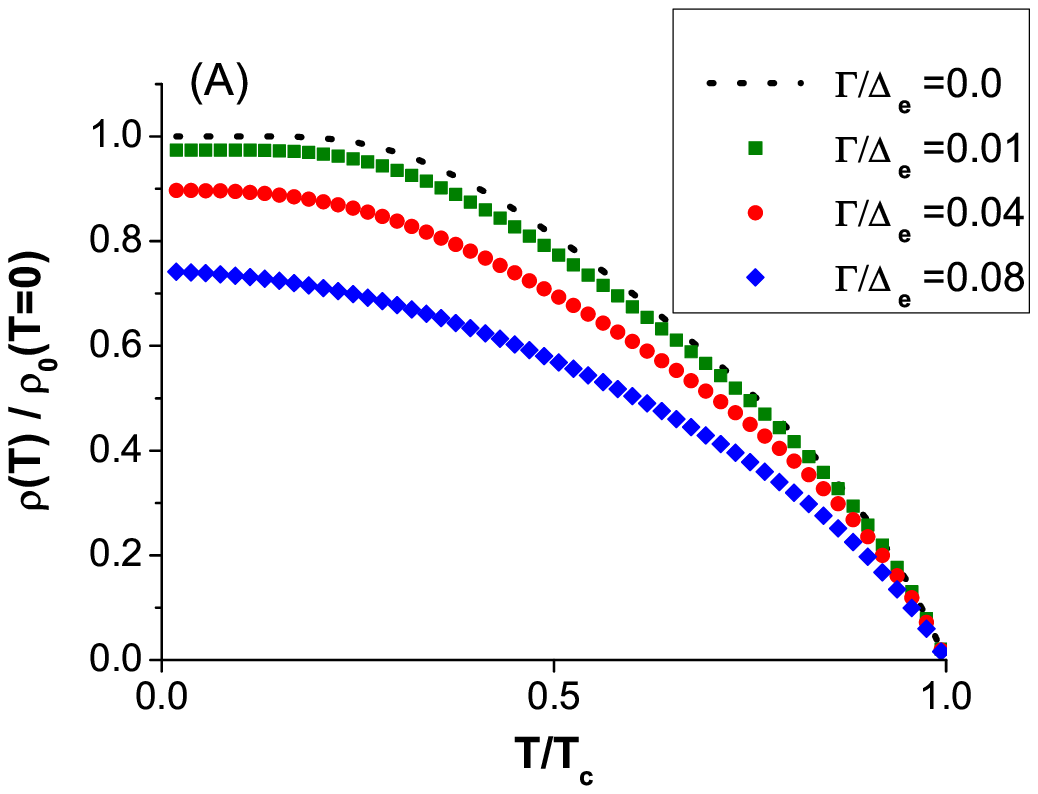}
\onefigure[width=80mm]{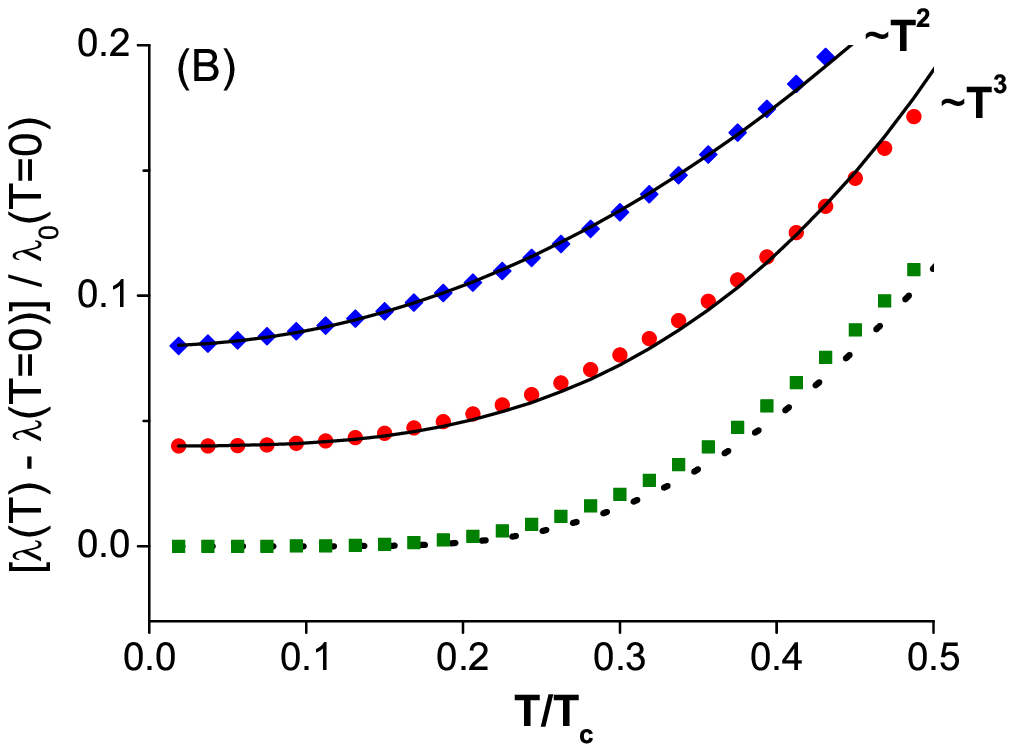}
\caption{(Color online) (a) Superfluid density $\rho_{s}(T)$ for
different impurity concentrations, $\Gamma / \Delta_e= 0.0, 0.01,
0.04, 0.08$, normalized by $\rho_s ^0(T=0)$ of the pure state. (b)
Corresponding penetration depth $\Delta \lambda(T)=
\lambda(T)-\lambda(T=0)$ normalized by $\lambda_0(T=0)$ of the
pure state. The data for $\Gamma / \Delta_e= 0.04, 0.08$ are offset
for clarity. The power law lines (sold black lines) of $T^2$ and
$T^3$ are shown for comparison.} \label{fig2}
\end{figure}

The above kernel is directly proportional to the superfluid
density $\rho_s (T)$ and 1/$\lambda_{L} ^2 (T)$ in the London
limit. We use this formula for numerical calculations. Before we
show the numerical results, we would like to gain analytic
understanding about the temperature dependence of the superfluid
density $\rho_s (T)$ of the $\pm$s-wave state when the DOS is
modified by impurity scattering as shown in Fig.\ref{fig1}. To
focus on the temperature dependence, we separate only the
temperature dependent part from Eq.(\ref{eq.8}) as

\begin{equation}
\label{eq.4}
\delta K(T)= -\frac{2 e^2}{c} \sum_{a=h,e} N_{a} \left \langle
v_{a \|} ^2 \int^{\infty} _{0} d\omega f(\omega) {\rm Re}
\frac{\tilde{\Delta}_{a}^2}{(\tilde{\omega}^2 -
\tilde{\Delta}_{a}^2)^{3/2}} \right \rangle, \label{eq.9}
\end{equation}

\noindent where $f(\omega)$ is Fermi-Dirac distribution function
and $\tilde{\omega}$ and $\tilde{\Delta}_{a} (\omega)$ are the
real frequency quantities obtained from $\tilde{\omega}_n$ and
$\tilde{\Delta}_{a} (\omega_n)$ by analytic continuation
($\omega_n \rightarrow \omega + i \eta$). As explained before, for
small impurity concentrations, $\tilde{\Delta}_{a} \approx
\Delta_{a}$ and $\tilde{\Delta}_{a}$ doesn't develop any peculiar
low energy structure, therefore we focus on the behavior of
$\tilde{\omega} = \omega + \Sigma^0 _{tot}(\omega)$ at low
energies. As can be seen in Fig.\ref{fig1}(b), for $\Gamma_{imp} >
\Gamma_{crit}$ ($\Gamma_{crit} =0.04 \Delta_e$), we can
approximate $\tilde{\omega} \approx a \omega + i \beta \omega + i
\gamma ...$ for low frequencies. Here $\gamma$ is the typical
impurity induced constant damping. Unusual part is the term $i
\beta \omega$ which gives a linear-in-$\omega$ damping that is a
unique feature of the $\pm$s-gap state with impurity scattering of
a strong coupling limit (unitary limit). With this form of
$\tilde{\omega}$, we can extract the temperature dependence of
$\delta K(T)$ using extended Sommerfeld expansion
\cite{Sommerfeld} at low temperatures.
After straightforward calculation, we obtain the following result.

\ba \label{eq.10}\delta K_a (T) & \approx & -\frac{2 e^2 N_a v_{a
\|} ^2}{c} \times \Biggl [\frac{\pi^2 \Delta_a ^2 a \gamma }{4
(\gamma^2+\Delta_a^2)^{5/2}} ~T^2  \\  \nonumber  &+& \frac{4
\Delta_a^2 a \beta }{(\gamma^2+\Delta_a^2)^{5/2}} \Biggl (1-
\frac{5 \gamma^2}{(\gamma^2+\Delta_a^2)} \Biggr)(1.35231) ~T^3
\Biggr] ...  \ea

\noindent As expected from a power counting, the constant damping
term $\gamma$ produces $T^2$ decrease and the linear-in-$\omega$
damping term $\beta \omega$ produces $T^3$ decrease of superfluid
density $\rho_s (T)$. The $T^2$ decrease of $\rho_s (T)$ due to a
constant damping is the well known correction due to the strong
coupling impurity scattering in the d-wave high-$T_c$ cuprates
\cite{Hirschfeld}. What is unusual is the $T^3$ correction due to
the linear-in-$\omega$ damping. This linear-in-$\omega$ damping is
the unique feature of the $\pm$s-wave SC with strong coupling
(unitary limit) impurity scattering and it modifies otherwise
fully opened DOS into a V-shape DOS. This V-shape DOS results in
the same thermodynamic behaviors as in a d-wave SC such as in
specific heat and NMR 1/$T_1$ relaxation rate, etc. However, when
transport property like the superfluidity density is calculated,
specific vertex enters combined with the coherent factors and a
naive picture based on the DOS fails.

With this result we have now a complete evolution of the
temperature dependence of the superfluid density $\rho_s(T)$ for
the $\pm$s-wave SC with strong coupling impurities. For a clean
sample ($\Gamma_{imp} < \Gamma_{crit}$), it will show an
exponentially flat dependence at low temperatures. With increasing
impurity concentration, this flat temperature region shrinks. At
the critical impurity concentration ($\Gamma_{imp}
=\Gamma_{crit}$), it will display the $T^3$ behavior. And finally
for $\Gamma_{imp} > \Gamma_{crit}$, the $T^2$ behavior will
smoothly dominate over the $T^3$ dependence. Therefore, for a
finite window of impurity concentrations of $\Gamma_{imp} >
\Gamma_{crit}$, experiments would display a power law $\propto
T^{\alpha}$ with $2 < \alpha < 3$. This evolution of temperature
dependence with impurity concentration is consistent with the data
of $M$-1111 ($M$=Pr, Nd, Sm) \cite{1111a,1111b,1111c} (flat),
(Ba,K)Fe$_2$As$_2$ \cite{Ba-122} (flat), and Ba(Fe,Co)$_2$As$_2$
\cite{BaCo-122-a,BaCo-122-b} ($\propto T^{2-2.5}$), except for the
data of LaFePO \cite{LaFePO} ($\propto T^{1.2}$).

In Fig.\ref{fig2}, we show the numerical results of $K(T)$ of
Eq.(\ref{eq.8}) using the same parameters as in Fig.\ref{fig1};
$K(T)$ is the same as the superfluid density $\rho_s(T)$ and
proportional to $1/\lambda_{L}^2 (T)$. For the temperature
dependence of the gaps $\Delta_{h,e}(T)$, we used a
phenomenological formula, $\Delta_{h,e}(T)=\Delta_{h,e}(T=0) \tanh
(1.74 \sqrt{T_{c}/T-1})$. The values of $\Delta _{h,e} / T_c$
determine the high temperature behavior below $T_c$. We chose $2
\Delta _{h} / T_c =3.0$ and $|\Delta_e| / |\Delta_h| = 2.5$ for
our calculations.

Figure \ref{fig2}(a) shows the superfluid density $\rho_s(T)$ over
$0 < T < T_c$ for different impurity concentrations, $\Gamma /
\Delta_e= 0.0, 0.01, 0.04, 0.08$, normalized by $\rho_s ^0(T=0)$
of the pure state. We warn the fact that the temperature
dependence of $\rho_s(T)$ at high temperatures (say, $0.5<T / T_c
< 1$) does not reflect the low energy DOS nor the impurity effects
but it is mainly determined by the temperature dependence of
$\Delta _{h,e}(T)$.
$\rho_s (T)$ of the pure case ($\Gamma / \Delta_e= 0.0$) shows the
exponentially flat temperature dependence at low temperatures,
consistent with the fully gapped DOS (see Fig.\ref{fig1}(a)) of
the $\pm$s-wave state. With increased impurity concentration,
$\Gamma / \Delta_e= 0.01$, this full gap behavior still persists.
When impurity concentration becomes critical, i.e., $\Gamma /
\Delta_e= 0.04$, the DOS become a perfect V-shape down to zero
frequency and the exponentially flat region of $\rho_s(T)$
disappears. With further increase of impurity concentration, for
$\Gamma / \Delta_e= 0.08$, the DOS is not only a V-shape but it
also obtains a finite DOS at zero frequency $N(0)=const.$, and
$\rho_s(T)$ becomes even more smooth. To see more clearly the
power laws of the low temperature behavior, we plot, in
Fig.\ref{fig2}(b), the temperature dependent part of the
penetration depth $\Delta \lambda(T)= \lambda(T)-\lambda(T=0)$
normalized by $\lambda_0(T=0)$ of the pure state. The power law
fittings (black solid lines in Fig.\ref{fig2}(b)) clearly confirm
that $\Delta \lambda(T) \propto T^3$ at the critical impurity
concentration $\Gamma_{imp}=\Gamma_{crit}=0.04 \Delta_e$ and
$\Delta \lambda(T) \propto T^2$ for higher concentration
$\Gamma_{imp}=0.08 \Delta_e$, as we have shown with an analytic
analysis of Eq.(\ref{eq.9}).

\section{Thermodynamic DOS and transport DOS} Switching gears, we
would like to draw attention to the more general aspect of DOS,
which was clearly demonstrated in this paper; namely the fact that
{\it the DOS measured by thermodynamic properties and the DOS
measured by transport properties can in general be different}. The
DOS of the $\pm$s-wave state with a critical impurity
concentration looks a V-shape as shown in Fig.\ref{fig1}(a). Here
the DOS is the distribution of the quasiparticle energy
eigenstates and this distribution can be measured by thermodynamic
properties such as specific heat, photoemission, and
spin-lattice-relaxation rate $1/T_1$, etc. Therefore, this V-shape
DOS is thermodynamically indistinguishable from the V-shape DOS of
a pure d-wave SC state. As mentioned, however,
angle-resolved-photoemission spectroscopy can still distinguish
the presence or absence of the angular (or Fermi surface)
anisotropy of the quasiparticle DOS
\cite{photoemission-s1,photoemission-s2,photoemission-s3}.

On the other hand, the superfluidity density is a transport
quantity defined as

\ba \label{eq.11}
\vec{J}(r) &=& -\frac{c}{4 \pi} K(T)
\vec{A}(r) = -\frac{c}{4 \pi \lambda^2(T)} \vec{A}(r) \nonumber \\
&=& -\frac{e^2}{mc} \rho(T) \vec{A}(r) \ea
\noindent in the local
limit (London limit). In the non-interacting case (no further
quasiparticle scattering after diagonalization of pairing
interactions), the BCS theory give a simple formula of the kernel
to the vector potential $\vec{A}$ as \cite{Tinkham}

\ba \label{eq.12}
K(T) = \frac{1}{\lambda_L ^2} \Big[1 -2
\int_{\Delta} ^{\infty} d \omega \Big(- \frac{\partial f}{\partial
\omega} \Big) Re \frac{\omega}{(\omega^2 - \Delta^2)^{1/2}} \Big].
\ea

The second term is apparently proportional to the thermodynamic
DOS $Re \frac{\omega}{(\omega^2 - \Delta^2)^{1/2}}$ and when
$\tilde{\omega}=\omega + i\eta$ the general expression of the
kernel (Eq.(\ref{eq.9})) can be converted into Eq.(\ref{eq.12}).
However, once the quasiparticle obtains a finite lifetime due to
interactions, Eq.(\ref{eq.9}) and Eq.(\ref{eq.12}) can not be
interchangeable and we have to use Eq.(\ref{eq.9}) to calculate
the superfluid density and penetration depth. Even for the
simplest case of a constant damping $\tilde{\omega}=\omega + i
\gamma$, using Eq.(\ref{eq.12}) would give $\delta \lambda(T)
\propto -const.$, but the correct temperature dependence due to a
constant damping is $\delta \lambda(T) \propto -\gamma T^2$ as
well known in the case of high-Tc cuprates \cite{Hirschfeld} and
reproduced in the current paper. When $\tilde{\omega}=\omega + i
\beta \omega$, the difference is even more interesting;
Eq.(\ref{eq.12}) predicts $\delta \lambda(T) \propto -\beta T$ but
the correct behavior from Eq.(\ref{eq.9}) produces $\delta
\lambda(T) \propto -\beta T^3$. We emphasize that this difference
of the DOS depending on the different measurements should be
carefully considered for interpretations of various experiments of
Fe pnitide SCs.

\section{Conclusion} We have calculated the superfluid density
$\rho_s(T)$ and penetration depth $\lambda(T)$ of the $\pm$s-wave
state with impurities of the strong coupling limit. We showed,
both by analytic analysis and by numerical calculations, that the
temperature dependence of $\rho_s(T)$ and $\lambda(T)$ at low
temperatures continuously evolves in a sequence of the forms:
exponentially flat $\rightarrow  \propto T^3 \rightarrow \propto
T^2$ with increase of impurity concentration. This result
consistently explain the experimental data of $M$-1111 ($M$=Pr,
Nd, Sm) \cite{1111a,1111b,1111c} (flat), (Ba,K)Fe$_2$As$_2$
\cite{Ba-122} (flat), and Ba(Fe,Co)$_2$As$_2$
\cite{BaCo-122-a,BaCo-122-b} ($\propto T^{2-2.5}$).
The near $T-$linear behavior of LaFePO \cite{LaFePO} data can not
be explained by our theory. If the data of LaFePO \cite{LaFePO} is
indeed confirmed, it indicates that LaFePO may have a different
pairing symmetry than the other Fe pnictide SCs. Finally, we
noticed that Vorontsov et al. \cite{Chubukov}, have recently
studied the same problem as in this paper, and obtained a similar
result $\rho_s(T)\propto T^2$ for high concentration of impurities
but obtained a different result $\rho_s(T)\propto T^{1.6}$ for the
critical impurity concentration. We think that the origin of this
difference mainly arises from the fact that Vorontsov et al.
\cite{Chubukov} studied a weak coupling impurity scattering and we
studied the strong coupling impurity scattering. As a consequence,
the definition of the critical impurity concentration and the
manner that the SC state becomes gapless with impurities is
different in each studies.

\acknowledgments This work was supported by the KOSEF through the
Grants No. KRF-2007-521-C00081.


\begin{thebibliography}{0}

\bibitem{Kamihara}
    \Name{Y.Kamihara et al.}
    \REVIEW{J. Am. Chem. Soc.}
{130}{2008}{3296}.

\bibitem{org1}
\Name{G. F. Chen, Z. Li, D. Wu, G. Li, W.Z. Hu, J. Dong, P. Zheng,
J.L. Luo, N.L. Wang}     \REVIEW{Phys. Rev. Lett.}
{100}{2008}{247002}.

\bibitem{org2}
    \Name{X. H. Chen, T. Wu, G. Wu, R. Liu, H. Chen, and D. Fang}
    \REVIEW{Nature (London)} {453}{2008}{761}.


\bibitem{tunneling-d1}
    \Name{Y. Wang, L. Shan, L. Fang, P. Cheng, C. Ren, and H. Wen}
arXiv:0806.1986 (unpublished).

\bibitem{tunneling-d2}
    \Name{L. Shan, Y. Wang, X. Zhu, G. Mu, L. Fang, C. Ren, and H.
Wen}
    \REVIEW{Europhys. Letters} {83}{2008}{57004}.

\bibitem{photoemission-d}
\Name{T. Sato, S. Souma, K. Nakayama, K. Terashima, K. Sugawara,
T. Takahashi, Y. Kamihara, M. Hirano, H. Hosono} \REVIEW{J. Phys.
Soc. Jpn.} {77}{2008}{063708}.

\bibitem{T1a}
\Name{K. Matano, Z.A. Ren, X.L. Dong, L.L. Sun, Z.X. Zhao,
Guo-qing Zheng} \REVIEW{Europhys. Lett.} {83}{2008}{57001}.


\bibitem{T1b}
\Name{H.-J. Grafe, D. Paar, G. Lang, N. J. Curro, G. Behr, J.
Werner, J. Hamann-Borrero, C. Hess, N. Leps, R. Klingeler, B.
Buechner} \REVIEW{Phys. Rev. Lett.}{101}{2008}{047003}.


\bibitem{T1c}
\Name{H. Mukuda, N. Terasaki, H. Kinouchi, M. Yashima, Y. Kitaoka,
S. Suzuki, S. Miyasaka, S. Tajima, K. Miyazawa, P.M. Shirage, H.
Kito, H. Eisaki, A. Iyo} \REVIEW{J. Phys. Soc. Jpn.} {77}{2008}{
093704}.

\bibitem{T1d}
\Name{Y. Nakai, K. Ishida, Y. Kamihara, M. Hirano, and H. Hosono}
\REVIEW{J. Phys. Soc. Jpn.} {77} {2008}{073701.}

\bibitem{T1e}
\Name{S. Kawasaki, K. Shimada, G. F. Chen, J. L. Luo, N. L. Wang,
and Guo-qing Zheng} \REVIEW{Phys. Rev. B} {78}{2008}{220506}.


\bibitem{tunneling-s}
\Name{T. Y. Chen, Z. Tesanovic, R. H. Liu, X. H. Chen, and C. L.
Chien} \REVIEW{Nature (London)} {453}{2008}{1224}.

\bibitem{photoemission-s1}
\Name{H. Ding et al.} \REVIEW{Euro. Phys. Lett.}
{83}{2008}{47001}.


\bibitem{photoemission-s2}
\Name{T. Kondo et al.} \REVIEW{Phys. Rev. Lett.}
{101}{2008}{147003}.


\bibitem{photoemission-s3}
\Name{L. Wray, D. Qian, D. Hsieh, Y. Xia, L. Li, J.G. Checkelsky,
A. Pasupathy, K.K. Gomes, A.V. Fedorov, G.F. Chen, J.L. Luo, A.
Yazdani, N.P. Ong, N.L. Wang, M.Z. Hasan} arXiv:0808.2185
(unpublished).

\bibitem{C(T)}
\Name{G. Mu, X. Zhu, L. Fang, L. Shan, C. Ren, and H. Wen}
\REVIEW{Chin. Phys. Lett.}
 {25}{2008}{2221}.

\bibitem{1111a}
\Name{L. Malone, J.D. Fletcher, A. Serafin, A. Carrington, N.D.
Zhigadlo, Z. Bukowski, S. Katrych, J. Karpinski} arXiv:0806.3908
(unpublished).


\bibitem{1111b}
\Name{K. Hashimoto, T. Shibauchi, T. Kato, K. Ikada, R. Okazaki,
H. Shishido, M. Ishikado, H. Kito, A. Iyo, H. Eisaki, S. Shamoto,
Y. Matsuda} \REVIEW{Phys. Rev. Lett.} {102}{2009}{017002.}

\bibitem{1111c}
\Name{C. Martin, R. T. Gordon, M. A. Tanatar, M. D. Vannette, M.
E. Tillman, E. D. Mun, P. C. Canfield, V. G. Kogan, G. D.
Samolyuk, J. Schmalian, R. Prozorov} arXiv:0807.0876
(unpublished).

\bibitem{Ba-122}
\Name{K. Hashimoto, T. Shibauchi, S. Kasahara, K. Ikada, T. Kato,
R. Okazaki, C. J. van der Beek, M. Konczykowski, H. Takeya, K.
Hirata, T. Terashima, Y. Matsuda} arXiv: 0810.3506 (unpublished).

\bibitem{BaCo-122-a}
\Name{R. T. Gordon, C. Martin, H. Kim, N. Ni, M. A. Tanatar, J.
Schmalian, I. I. Mazin, S. L. Bud'ko, P. C. Canfield, R. Prozorov}
arXiv:0812.3683 (unpublished).

\bibitem{BaCo-122-b}
\Name{R. Prozorov, M. A. Tanatar, R. T. Gordon, C. Martin, H. Kim,
V. G. Kogan, N. Ni, M. E. Tillman, S. L. Bud'ko, P. C. Canfield}
arXiv:0901.3698 (unpublished).

\bibitem{LaFePO}
\Name{J. D. Fletcher, A. Serafin, L. Malone, J. Analytis, J-H Chu,
A.S. Erickson, I.R. Fisher, A. Carrington} arXiv:0812.3858
(unpublished).


\bibitem{Mazin}
\Name{I.I. Mazin, D.J. Singh, M.D. Johannes, M.H. Du} \REVIEW{
Phys. Rev. Lett.}{101}{2008}{057003.}

\bibitem{other-theory1}
\Name{K. Kuroki, S. Onari, R. Arita, H. Usui, Y. Tanaka, H.
Kontani, and H. Aoki } \REVIEW{Phys. Rev. Lett. } {101}{2008}{
087004.}


\bibitem{other-theory2}
\Name{M. M. Korshunov and I. Eremin} \REVIEW{Phys. Rev. B}
{78}{2008}{140509.}


\bibitem{other-theory3}
\Name{F. Wang, H. Zhai, Y. Ran, A. Vishwanath, Dung-Hai Lee}
\REVIEW{Phys. Rev. Lett.}{102}{2009}{047005.}

\bibitem{Bang-model}
\Name{Y. Bang and H.-Y. Choi} \REVIEW{Phys. Rev. B}
{78}{2008}{134523.}

\bibitem{Bang-imp}
\Name{Y. Bang, H.-Y. Choi, and H. Won} \REVIEW{Phys. Rev. B}
{79}{2009}{054529.}

\bibitem{theory-T1a}
\Name{D. Parker, O.V. Dolgov, M.M. Korshunov, A.A. Golubov, I.I.
Mazin } \REVIEW{Phys. Rev. B } {78}{2008}{134524.}

\bibitem{theory-T1b}
\Name{A.V. Chubukov, D.V. Efremov, I. Eremin} \REVIEW{Phys. Rev.
B} {78}{2008}{134512.}

\bibitem{theory-T1c}
\Name{M. M. Parish, J. Hu, B. A. Bernevig} \REVIEW{Phys. Rev. B}
{78}{2008}{144514.}


\bibitem{T-mtx-1}
\Name{P. J. Hirschfeld, P. Woelfle, and D. Einzel} \REVIEW{Phys.
Rev. B }{37}{1988}{83.}

\bibitem{T-mtx-2}
\Name{A. V. Balatsky, I. Vekhter, and J.-X. Zhu} \REVIEW{Rev. Mod.
Phys.}{78}{2006}{373}, and see more references therein.


\bibitem{Sommerfeld}
Because the integration domain is only over the positive
frequencies as $\int_0 ^{\infty} d \omega$, we include the
$\omega^3$ term whereas the odd power terms are assumed to
trivially vanish in the ordinary Sommerfeld expansion.


\bibitem{Hirschfeld}
\Name{P.J. Hirschfeld and N. Goldenfeld} \REVIEW{Phys. Rev. B}
{48}{1993}{4219.}

\bibitem{Tinkham}
\Name{M. Tinkham} \Book{in Chap. 3, Introduction to
superconductivity, 2nd edition}
 \Publ{McGraw-Hill Inc., New York}
 \Year{1996.}

\bibitem{Chubukov}
\Name{A.B. Vorontsov, M.G. Vavilov, A.V. Chubukov} arXiv:0901.0719
(unpublished).

\end{thebibliography}
\end{document}